\documentclass{article}

\usepackage{fullpage}
\usepackage{amsthm}
\usepackage{amsmath}
\usepackage{amsfonts}
\usepackage{hyperref}
\usepackage{bm}
\usepackage{graphicx}

\newcommand{\Ket}[1]{\ensuremath{\left \vert #1 \right \rangle}}

\newcommand{\BraKet}[2]{\ensuremath{\left \langle #1 \middle \vert #2
    \right \rangle}}
\newcommand{\QProb}[2]{\left \vert \BraKet{#1}{#2} \right \vert^2}
\newcommand{\Tr}[2][]{\ensuremath{\mathrm{Tr}_{#1} \left ( #2 \right )}}

\title{Time Symmetric Quantum Theory Without Retrocausality?  A Reply
  to Tim Maudlin}

\author{Matthew S. Leifer \\ Institute for Quantum Studies \& Schmid
  College of Science and Technology \\ Chapman University, Orange, CA
  92866, USA}

\date{\today}

\begin{document}

\maketitle

\begin{abstract}
  In \href{https://arxiv.org/abs/1707.08641}{arXiv:1707.08641}, Tim
  Maudlin claims to construct a counterexample to the result of
  \href{http://rspa.royalsocietypublishing.org/content/473/2202/20160607}{Proc. Roy. Soc. A
    vol. 473, iss. 2202, 2017}
  (\href{https://arxiv.org/abs/1607.07871}{arXiv:1607.07871}), in
  which it was shown that no realist model satisfying a certain notion
  of time-symmetry (in addition to three other assumptions) can
  reproduce the predictions of quantum theory without retrocausality
  (influences travelling backwards in time).  In this comment, I explain
  why Maudlin's model is not a counterexample because it does not
  satisfy our time-symmetry assumption.  I also explain why Maudlin's
  claim that one of the Lemmas we used in our proof is incorrect is
  wrong.
\end{abstract}

\section{Introduction}

\begin{figure}[!htb]
  \centering
  \includegraphics[scale=0.7]{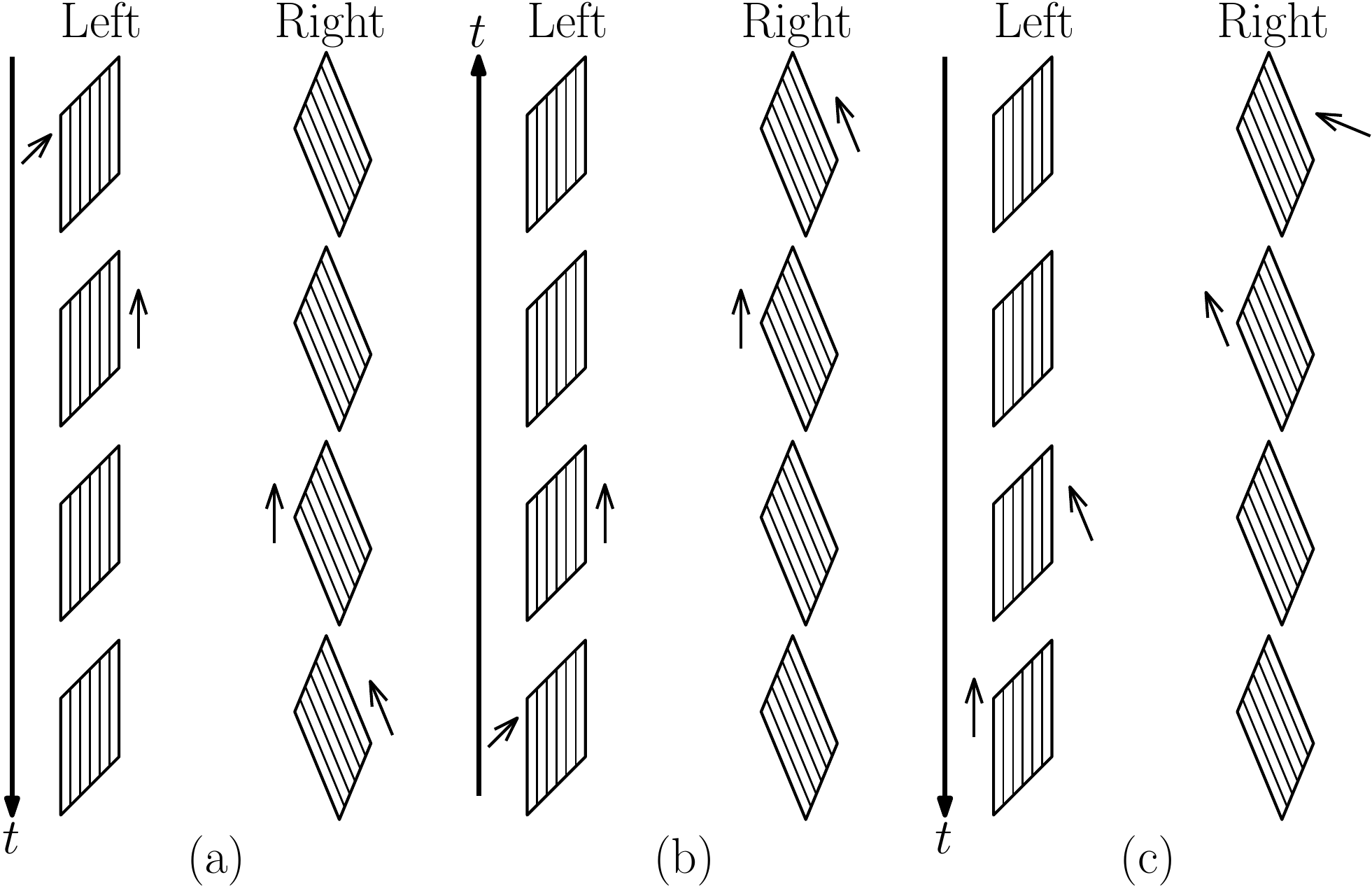}
  \caption{\label{fig1}Time reversal in polarization experiments.  (a)
    An experiment where the photon travels from the left to the right
    polaroid filter.  Time runs from top to bottom.  (b) The exact
    same experiment, but with the direction of time reversed to run
    bottom to top. (c) An experiment where the photon travels from the
    right polarizer to the left one, with time running top to bottom.
    The evolution of the polarization vector is not the same as in
    (b).}
\end{figure}

In his book, Time's Arrow and Archimedes' Point \cite{Price1996}, Huw
Price argued that the ``standard interpretation''\footnote{A loaded
  term, because there is widespread disagreement about what the
  ``standard interpretation'' actually is, but here we mean something
  like the interpretation contained in the books of Dirac
  \cite{Dirac1958} and von Neumann \cite{Neumann1955}, which endorses
  the collapse of the wavefunction and the eigenvalue-eigenstate link.
  This renders the quantum state an ontic property of the system that
  changes discontinuously upon measurement.} of quantum theory is time
asymmetric, and that retrocausality is required to restore time
symmetry.  The argument runs as follows:

Consider a photon travelling between two crossed polarizers, let's
call them the left polarizer and the right polarizer (see
fig.~\ref{fig1}(a)).  As the photon exits the left polarizer, its
polarization vector is aligned with the direction of that polarizer.
It then encounters the right polarizer, still with its polarization
vector aligned with the left polarizer, and either passes through or
not with the probabilities ascribed by quantum theory.  Suppose that,
in this run of the experiment, it passes, and hence exits with a
polarization vector aligned with the right polarizer.

If we run this description backwards in time (see fig.~\ref{fig1}(b)),
we find that the photon starts with its polarization aligned with the
right polarizer, but as soon as it passes through the right polarizer
its polarization immediately jumps to be aligned with that of the left
polarizer.  It then carries on towards the left polarizer and passes
through it.  If time symmetry holds then this ought to be a possible
description of a forwards-in-time experiment, but that would involve
retrocausality as the photon seems to anticipate the direction of the
left polarizer, which could be chosen while the photon is still in
flight between the two.

Of course, this is not at all what happens in the actual quantum
description of the time reversed experiment (see fig.~\ref{fig1}(c)).
Instead, it is just like the description of the original experiment
except with the roles of the left and right polarizers interchanged.
In this description, the photon polarization is aligned with the right
polarizer during its flight between the two polarizers, so there is no
retrocausality, but we seem to have lost time symmetry, i.e.\ running
the description of the first experiment backwards in time does not
give us a valid description of a forwards-in-time experiment.

This argument is subject to two obvious objections.  Firstly, it is
only unitary time evolution according to the Schr{\"o}dinger equation
that is time symmetric in quantum theory.  To get a time-symmetric
description, we ought to treat the whole experiment, including the
interaction with the polarizers, according to unitary quantum
mechanics.  This is easiest to do if we use polarizing beam-splitters
rather than polaroid filters for the polarization measurements.  Then,
we see that, in the first experiment, the photon exits the right
polarzing beam-splitter in a superposition of being in the two output
ports.  If we run this description backwards in time then the photon
enters the right beam-splitter in a superposition of the two ports,
and these interfere to yield a polarization direction that happens to
be the same as the orientation of the left beam-spliiter.  This is a
valid description of a forwards-in-time experiment, so time symmetry
is restored.  There is also no retrocausality here as the photon
polarization is determined by the initial amplitudes at the input
ports of the right polarizer, and not by the orientation of the left
polarizer, which can still be varied independently.

The error, according to this objection, is misapplication of the
measurement postulates in our first attempt to describe time reversal.
What happens during measurement is, of course, controversial, but in
the conventional formalism the collapse of the quantum state at least
looks like a time asymmetric process.  It is only unitary evolution
that is claimed to be time reversible in the standard interpretation
of quantum theory.

However, this objection ignores the fact that there is a form of time
symmetry that obtains even if we do apply the measurement postulates.
Returning to our first attempt to describe time reversal, let
$\Ket{\psi_l}$ be the polarization state of a photon that passes the
left polaroid filter with certainty and let $\Ket{\psi_r}$ be the
polarization state of a photon that passes the right polaroid filter
with certainty.  Then, the probability of passing the right polarizer
given that the photon has passed the left polarizer is
$\QProb{\psi_r}{\psi_l}$.  In the time reversed experiment, the
probability of passing the left polarizer given that the photon passed
the right polarizer is $\QProb{\psi_l}{\psi_r} =
\QProb{\psi_r}{\psi_l}$, i.e.\ the exact same probability.

This time symmetry can be generalized.  Consider $n$ polaroid filters
placed in sequence, with $\Ket{\psi_j}$ being the state that passes
the $j^{\text{th}}$ filter with certainty.  Then, the joint
probability of passing filters $2,3,\ldots,n$ given that the first
filter was passed is $\prod_{j=1}^{n-1} \QProb{\psi_{j+1}}{\psi_j}$,
and in the time reversed experiment the probability of passing filters
$n-1,n-2,\ldots ,1$ given that the $n^{\text{th}}$ filter was passed
is also $\prod_{j=1}^{n-1} \QProb{\psi_{j+1}}{\psi_j}$.

Generalizing even further, we find that \emph{any} sequence of
measurements interspersed by dynamics can be described either in the
conventional predictive formalism, in which states are evolved
forwards in time, or in a retrodictive formalism, in which states are
evolved backwards in time (see \cite{Leifer2013a} and references
therein).  Because the predictive and retrodictive formalisms are
mathematically identical, we conclude that for any forwards-in-time
experiment consisting of measurements interspersed with dynamics there
is another forwards-in-time experiment that predicts the same
probabilities as the first experiment running backwards in time.

Therefore, it is possible that the apparent time asymmetry of quantum
measurements is just an artifact of the way they are described in the
conventional formalism.  There exist other formalisms, such as the two
state vector formalism of Aharonov et.\ al. \cite{Aharonov2008}, the
consistent/decoherent histories formalism \cite{Griffiths2002}, and
the path integral formalism \cite{Feynman1965}, in which the symmetry
just described is more explicit.  These formalisms all either have the
feature that the state at the present time depends on the future as
well as the past, or that the formalism only refers to spacetime as a
whole and does not assign states at a given instant of time.  This
seems to indicate some form of retrocausality, but one should be
cautious because these are just mathematical formalisms, which can be
interpreted in a variety of ways, rather than clean statements about
the nature of reality.

In any case, what is certain is that there is a kind of time symmetry
in quantum theory that includes measurements, and that is different
from the conventional time reversal of unitary dynamics.  On this
notion, Price is correct in his description of time reversal, and his
argument that time symmetry requires retrocausality holds more water.
In his paper \cite{Price2012}, Price presents a modified and more
rigorous version of his argument, identifying more clearly the
circumstances under which it holds.  Our paper \cite{Leifer2017}
follows on from this, and makes the argument even more general.

This brings us to the second objection, which applies if one takes the
view that the quantum state ought to be thought of as an epistemic
state (state or knowledge, information, or belief) akin to a
probability distribution, as opposed to an ontic state (state of
reality).  In a time symmetric theory, we expect the ontic description
of an experiment to be reversible in time, but there is no reason why
an epistemic description should be.  This is because the knowledge
acquired by an observer also has a time direction to it, i.e.\ the
observer assigns the photon a polarization state with the same
orientation as a polarizer it has just passed through because the
observer \emph{knows} it has just passed through that polarizer.  The
apparent discontinuous jump of the photon polarization to the
orientation of the left polarizer in the reversed description occurs
because it is still the description assigned by an observer who knows
that the photon will pass through left polarizer.  If we also reverse
the direction of the observer's information acquisition, so that they
know the photon has passed the right polarizer in the reversed
experiment but not whether it will pass the left one, then we get a
different description, which is in fact just the usual quantum
description of the time reversed experiment.  In fact, all the
qualitative features used in Price's argument occur within Spekkens'
toy model \cite{Spekkens2007}, which is time symmetric and in which
the analog of a quantum state is a probability distribution.  This is
discussed in detail in our paper.

The main contribution of our argument is to show that, in order to
reproduce the predictions of quantum mechanics, you need
retrocausality to have the kind of time symmetry discussed above, even
if the quantum state is epistemic.

The reason why I have described this history and motivation in detail
is that Maudlin's purported counter example to our theorem
\cite{Maudlin2017} \emph{is} just what I have been calling the
standard interpretation of quantum theory.  His state of reality is
just the usual forwards evolving quantum state vector, and so his
model is already subject to Price's arguments, without even having to
consider ours.  It is not time symmetric in the sense that includes
measurements, just for the reasons given above.

For those still not convinced, in the remainder of this paper, I review
our assumptions, show that Maudlin's model fails to satisfy them, and
show that Maudlin's criticism of Lemma VIII.2 in our paper is incorrect.

\section{Our Assumptions}

\subsection{Operational Framework}

\begin{figure}[!htb]
  \centering
  \includegraphics[scale=0.7]{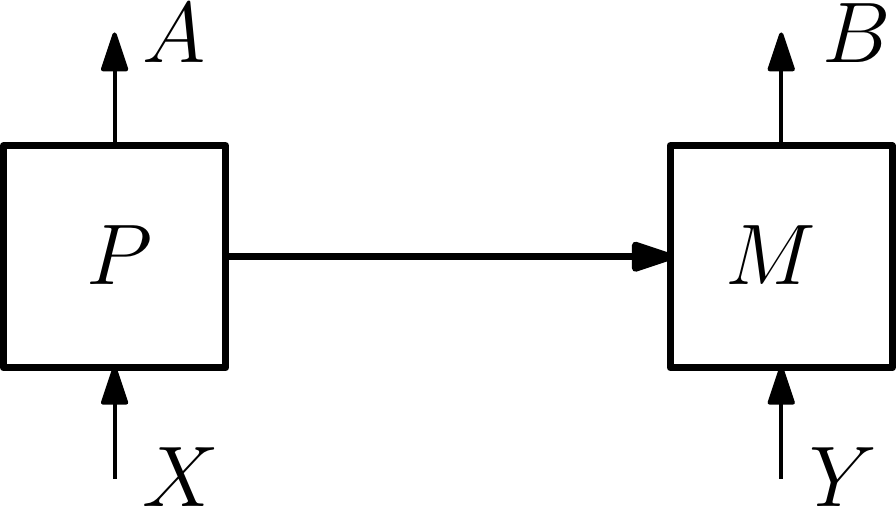}
  \caption{\label{fig2}Schematic of an experiment.}
\end{figure}

In \cite{Leifer2017}, we discussed a class of experiments, which are
described operationally as a preparation procedure $P$, followed by a
transformation $T$, followed by a measurement $M$.  Maudlin correctly
notes that the transformation $T$ does not play any role in our main
argument.  We included it in our paper to get a cleaner relationship
between quantum experiments that violate our assumptions and states
that violate Bell inequalities in a spacelike Bell test, as discussed
in \S VIII.C of \cite{Leifer2017}.  For present purposes, we can eliminate
$T$ and just discus prepare-and-measure experiments, as depicted in
fig.~\ref{fig2}.

The preparation procedure has an input $X$ and an output $A$, and
similarly the measurement $M$ has an input $Y$ and an output $B$.  The
outputs $A$ and $B$ are random variables that take values in (finite)
sets $\Omega_A$ and $\Omega_B$ respectively\footnote{I am being more
  explicit about the fact that the outputs are random variables than
  in the original paper to make the meaning of our \text{Time Symmetry}
  assumption clearer.}, and the inputs are variables\footnote{These are
  not random variables as they are chosen by the experimenter and
  always appear to the right of the conditional.} $X$ and $Y$ that
take values in sets $\Omega_X$ and $\Omega_Y$ respectively.  A pair
$(P,M)$ is called an experiment if $P$ and $M$ can be performed
sequentially in time on the same system.  For each experiment $(P,M)$,
an operational theory provides predictions of the probabilities
$p_{PM}(A = a,B = b|X = x,Y = y)$ for any $a \in \Omega_A$, $b \in
\Omega_B$, $x \in \Omega_X$ and $y \in \Omega_Y$.

In quantum theory, a preparation $P$ is described by a set
$\{\rho_{a|x}\}$ of unnormalized density operators, such that
$\sum_{a\in\Omega_A} \rho_{a|x}$ is a normalized density operator for
every $x\in\Omega_X$, and a measurement $M$ by a set of Positive
Operator Valued Measures (POVMs) $\{E_{b|y}\}$, i.e.\ for each
$y\in\Omega_Y$, $\{E_{b|y}\}_{b\in\Omega_B}$ is a set of positive
operators such that $\sum_{b \in \Omega_B} E_{b|y} = I$, where $I$ is
the identity operator.  The probabilities predicted for an experiment
are then
\[p_{PM}(A = a,B = b|X=x,Y=y) = \Tr{E_{b|y}\rho_{a|x}}.\]

\subsection{Ontological Framework}

So far, we have just described the aspects of experiments that can be
directly manipulated or observed in the lab, but we are interested in
investigating realist accounts in which the system has some objective
physical properties that are probed by the experiment.

We made four assumptions about such models in our paper:
\textbf{(Single World) Realism}, \textbf{Free Choice},
$\bm{\lambda}$\textbf{-mediation}, and \textbf{No Retrocausality}.
These are equivalent to assuming that the operational theory has an
\emph{ontological model}, which is a standard framework for discussing
realist models of quantum theory \cite{Harrigan2010}.  The reason we
broke this down into four assumptions is because we were interested in
investigating the role of \textbf{No Retrocausality}, which is baked
into the assumptions of an ontological model.  Nonetheless, it is
equivalent to just assume that our operational theory has an
ontological model, so that is what we will do here.

The ontological models framework says that a system has some objective
physical properties, described by its \emph{ontic state} $L$.  This is
again a random variable that takes values in a set $\Lambda$ called
the \emph{ontic state space}.  For simplicity, we shall assume that
$\Lambda$ is finite in our formal analysis, but the continuum and
measure-theoretic generalizations are straightforward.  The ontic
state $L$ is responsible for mediating any correlations between the
preparation and measurement variables that we observe in the lab.  For
every experiment $(P,M)$, an ontological model assigns probabilities
$p_{PM}(A = a,B = b,L = \lambda|X = x,Y = y)$ of the form
\begin{multline}\label{OM}
  p_{PM}(A = a,B = b,L = \lambda|X = x,Y = y) \\ =
  p_{M}(B = b|L = \lambda,Y = y)p_P(L = \lambda|A = a,X = x)p_P(A =
  a|X = x).\end{multline}
These reproduce the operational predictions if
\[\sum_{\lambda\in\Lambda} p_{PM}(A = a,B = b,L = \lambda|X = x,Y = y) =
  p_{PM}(A = a,B = b|X = x,Y = y),\]
which we shall always assume.

\begin{figure}[!htb]
  \centering
  \includegraphics[scale=0.7]{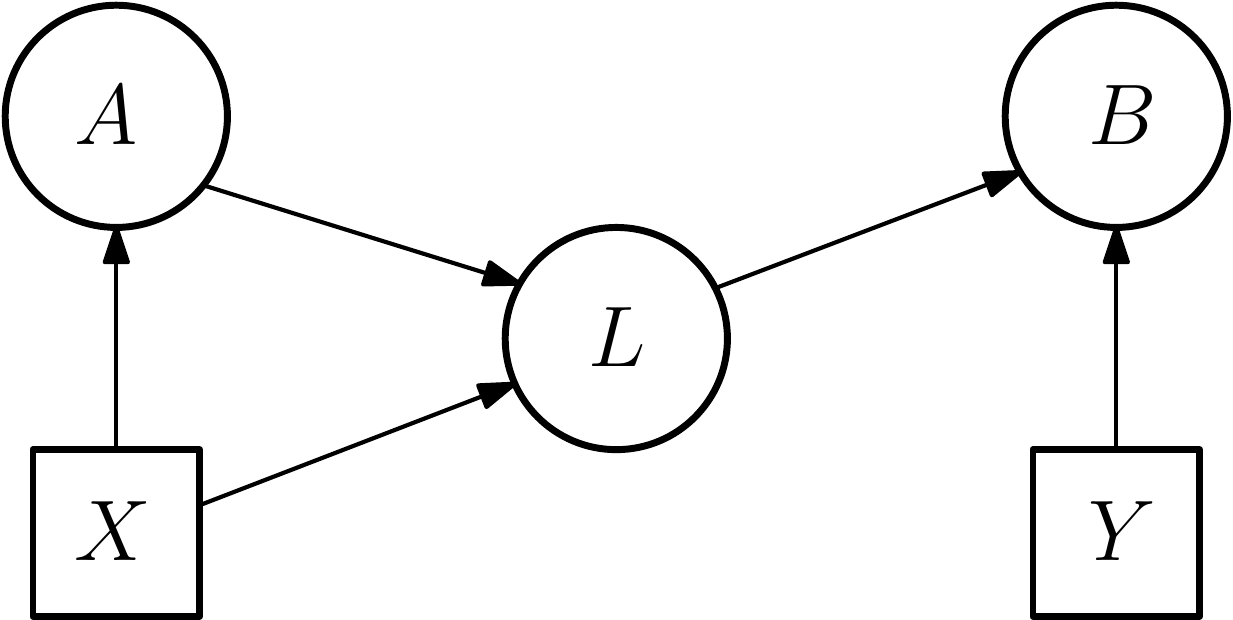}
  \caption{\label{fig3}Influence diagram showing the causal structure of an
    ontological model of an experiment.}
\end{figure}

Eq.~\eqref{OM} says that the observed probabilities are accounted for as
follows.  First, the experimenter chooses $X = x$ and the preparation
device responds by outputting $A = a$ with probability $p_P(A = a|X = x)$
and outputting $L = \lambda$ with probability
$p_P(L = \lambda|A = a,X = x)$.  $L$ is then transmitted to the
measurement device, the experimenter picks $Y = y$, and the measurement
device outputs $B = b$ with probability $p_{M}(B = b|L = \lambda,Y = y)$.
Importantly, this does not depend on $P$, $X$ or $A$ as we want the
properties of the system, i.e.\ the ontic state $L$, to mediate the
correlations between preparation and measurement.  The causal
structure of this type of model is depicted in fig.~\ref{fig3}.

\subsection{Time Symmetry}

The most controversial assumption in our argument is our \textbf{Time
  Symmetry} assumption because it is intended to capture the time
symmetry under sequential measurements discussed in the introduction,
rather than the conventional notion of time symmetry under unitary
evolution.  We start by defining what we mean by this symmetry for a
prepare-and-measure experiment.

An experiment $(P,M)$ has an \emph{operational time reverse} if there
exists another experiment $(P',M')$ (we use primes to denote the
variables associated with $(P',M')$) such that
$\Omega_{A'} = \Omega_B$, $\Omega_{B'} = \Omega_A$,
$\Omega_{X'} = \Omega_Y$, $\Omega_{Y'} = \Omega_X$, and
\begin{equation}
  \label{OpTR}
  p_{PM}(A = a,B = b|X = x,Y = y)=p_{P'M'}(A' = b,B' = a|X' = y,Y' =
  x).
\end{equation}
If every experiment in an operational theory has an operational time
reverse then the theory is \emph{operationally time symmetric}.

Note that an experiment can be its own operational time reverse, in
which case we have
\[p_{PM}(A = a,B = b|X = x,Y = y)=p_{PM}(A = b,B = a|X = y,Y = x).\]

We do not expect theories to be operationally time symmetric in
general because of the possibility of sending signals into the future
but not into the past.  That is, we expect that for all experiments
$(P,M)$, $p_{PM}(B = b|X = x,Y = y) \neq p_{PM}(B = b|Y = y)$ but
$p_{PM}(A = a|X = x,Y = y) = p_{PM}(A = a|X = x)$.  However, for any
operational theory, we can always consider the set of
\emph{no-signalling preparations}, i.e.\ the set of preparations $P$
such that $p_{PM}(B = b|X = x,Y = y) = p_{PM}(B = b|Y = y)$ happens to
be satisfied for all experiments $(P,M)$ involving $P$. For some
theories the set of no-signalling preparations might be empty, but it
can always be constructed.  The \emph{no-signalling sector} of an
operational theory is the operational theory that results from
restricting the preparations to be no-signalling.

In \cite{Leifer2017}, we showed that the no-signalling sector of
quantum theory is operationally time symmetric.  This also holds for
several other theories, including classical probability theory and the
Spekkens' toy theory \cite{Spekkens2007}.  Maudlin asserts that we
proved that the no-signalling sector of any operational theory is
operationally time-symmetric, but this is not the case.  It can fail
to hold for trivial reasons, e.g.\ if there is a preparation with
exactly two possible outcomes but all the measurements in the theory
have exactly three outcomes so no mapping from $P$ to $M'$ and $M$ to
$P'$ exists.  However, it can also fail for less trivial reasons.  It
is generically false for generalized probabilistic theories
\cite{Janotta2014}, in which the structure of the state space might be
very different from the structure of the set of \emph{effects} that
describe measurement outcomes.  Therefore, we claim that operational
time symmetry of the no-signalling sector is a significant symmetry of
quantum theory that is worth investigating.  If Maudlin's claim were
true then we would have no grounds for this.

We next follow the logic of Spekkens' noncontextuality
\cite{Spekkens2005} in asserting that this operational time symmetry
would be best explained if there were an analogous time symmetry at
the ontological level.  This is a form of Leibniz's principle of the
identity of indiscernibles: if two things look the same from the point
of view of all observations we can make on them then we should think
of them as actually being the same.  This motivates the following
definitions.

An experiment $(P,M)$ in an ontological model has an \emph{ontological
  time reverse} if there exists another experiment $(P',M')$ such that
$\Omega_{A'} = \Omega_B$, $\Omega_{B'} = \Omega_A$,
$\Omega_{X'} = \Omega_Y$, $\Omega_{Y'} = \Omega_X$, there exists a
one-to-one map $f:\Lambda \rightarrow \Lambda'$, and
\begin{equation}
  \label{ontTR}
  p_{PM}(A = a,B = b,L = \lambda|X = x,Y = y) = p_{P'M'}(A' = b,B' =
  a,L' = f[\lambda]|X' = y,Y' = x).
\end{equation}

The reason for including the function $f$ is that physical states
often have to be transformed when taking the time reverse, usually by
an involution.  For example, in Hamiltonian mechanics, the momentum is
reversed upon taking time reverses and, in quantum theory, the state
vector $\Ket{\psi}$ is mapped to its conjugate $\Ket{\psi^*}$ with
respect to some basis so we would need a nontrivial $f$ in a model
where the state vector is ontic.
 
Note that if $(P,M)$ and $(P',M')$ are ontological time reverses of
each other then they must also be operational time reverses, because
summing eq.~\eqref{ontTR} over $\lambda$ yields eq.~\eqref{OpTR}.
However, the converse is not necessarily true.

Our \textbf{Time Symmetry} assumption is then the assertion that if an
experiment has an operational time reverse then it should also have an
ontological time reverse. 

\section{The Beltrametti-Bugajski Model}

Before discussing Maudlin's model, it is useful to study a more
general model, known as the Beltrametti-Bugajski model
\cite{Beltrametti1995}.  Here we will only need the special case of a
spin-$1/2$ particle and orthonormal basis measurements.  Maudlin's
model is just this special case specialized further to only two
possible preparation and measurement directions, with a relabelling of
the inputs.  The reason it fails the \textbf{Time Symmetry} assumption
will be the same.

Consider the following preparation $P$ of a spin-$1/2$ particle.  A
direction is chosen, which we represent by a unit vector
$X = \vec{n}$.  Then, with probability $1/2$ the output $A = +1$ is
generated and with probability $1/2$ the output $A = -1$ is generated.
If the output is $+1$ then the system is prepared in the state
$\Ket{\vec{n},+1}$, i.e.\ spin-up in the direction $\vec{n}$ and if
$A=-1$ the system is prepared in the state $\Ket{\vec{n},-1}$, i.e.\
spin-down in the direction $\vec{n}$.  This is a no-signalling
preparation, as the ensemble average state over $A$ is always the
maximally mixed state, regardless of the direction $X$.

The measurement $M$ consists of choosing a direction $Y = \vec{m}$ and
measuring the system in the $\Ket{\vec{m},1+}, \Ket{\vec{m},-1}$ basis.
We set $B=+1$ if the outcome is $\Ket{\vec{m},+1}$ and $B=-1$ if the
outcome is $\Ket{\vec{m},-1}$.

For this experiment, quantum theory predicts the probabilities
\[p_{PM}(A = a,B = b|X = \vec{n},Y = \vec{m}) = \frac{1}{4} \left ( 1 + (ab) \vec{n}
    \cdot \vec{m} \right ).\]
This experiment is its own operational time reverse, i.e.
\[p_{PM}(A=a,B=b|X=\vec{n},Y=\vec{m}) = p_{PM}(A=b,B=a|X = \vec{m},Y=\vec{n}).\]

The Beltrametti-Bugajski model is really just a translation of the
orthodox Dirac-von Neumann interpretation into the ontological models
framework, i.e.\ the ontic state is identified with the quantum state
and the measurement probabilities are those given by quantum
mechanics.  

One way of implementing this is to choose the ontic state space to be
$\Lambda = S^2 \times \{+1,-1\}$, i.e.\ the Cartesian product of the
unit sphere with the two possible values $\pm 1$.  The ontic states
are of the form $\lambda = [\vec{\lambda}_1,\lambda_2]$ where
$\vec{\lambda}_1 \in S^2$ and $\lambda_2 = \pm 1$.  The probabilities
of the ontological model are specified as follows:
\begin{align*}
  p_P(A = a|X = \vec{n}) & = \frac{1}{2} \\
  p_P(L = [\vec{\lambda}_1,\lambda_2]|X = \vec{n},A = a) & =
                                                   \delta(\vec{\lambda}_1 - \vec{n}) \delta_{\lambda_2,a} \\
  p_M(B = b|L = [\vec{\lambda}_1,\lambda_2],Y = \vec{m}) & = \frac{1}{2} \left ( 1 +
                                             (\lambda_2 b)
                                             \vec{\lambda}_1 \cdot \vec{m}
                                             \right ),
\end{align*}
from which we can compute the probabilities
\begin{equation}
  \label{BBProbs}
  p_{PM}(A = a,B = b,L = [\vec{\lambda}_1,\lambda_2]|X = \vec{n},Y = \vec{m}) = \frac{1}{4}
  \delta(\vec{\lambda}_1 - \vec{n}) \delta_{\lambda_2,a} \left ( 1 +
                                             (\lambda_2 b)
                                             \vec{\lambda}_1 \cdot \vec{m}
                                           \right ).
\end{equation}
It is straightforward to see that this reproduces the quantum
predictions.

It is also easy to check that this model \emph{does not} satisfy the
\textbf{Time Symmetry} assumption.  Since this experiment is its own
operational time reverse, \textbf{Time Symmetry} says that we should
have
\begin{equation}
  \label{MaudTS}
  p_{PM}(A = a,B = b,L = [\vec{\lambda}_1,\lambda_2]|X = \vec{n},Y = \vec{m}) =
  p_{PM}(A = b,B = a,L = f[\vec{\lambda}_1,\lambda_2]|X = \vec{m},Y = \vec{n}).
\end{equation}
Define the function
$g(a,b,\vec{\lambda}_1,\lambda_2,\vec{n},\vec{m}) = p_{PM}(A = a,B =
b,L = [\vec{\lambda}_1,\lambda_2]|X = \vec{n},Y = \vec{m})$ and let
$g(\vec{\lambda}_1,\vec{n},\vec{m}) = \sum_{a,b,\lambda_2}
g(a,b,\vec{\lambda}_1,\lambda_2,\vec{n},\vec{m})$.  By direct
computation from eq.~\eqref{BBProbs}, we have
\begin{equation}
  \label{g1}
  g(\vec{\lambda_1},\vec{n},\vec{m}) = \delta(\vec{\lambda}_1 - \vec{n}).
\end{equation}

To apply eq.~\eqref{MaudTS}, we write
$f[\vec{\lambda}_1,\lambda_2] =
[\vec{f}_1(\vec{\lambda}_1,\lambda_2),f_2(\vec{\lambda_1},\lambda_2)]$,
where $\vec{f}_1(\vec{\lambda_1},\lambda_2) \in S^2$ and
$f_2(\vec{\lambda_1},\lambda_2) \in \{+1,-1\}$.  Then, combining
eqs.~\eqref{MaudTS} and \eqref{BBProbs} gives
\begin{equation}
  \label{g2}
  g(\vec{\lambda}_1,\vec{n},\vec{m}) =
  \delta(\vec{f}_1(\vec{\lambda}_1,\lambda_2) - \vec{m}).
\end{equation}
However, eqs.~\eqref{g1} and \eqref{g2} cannot possibly be the same
function because the first depends on $\vec{n}$ and not $\vec{m}$ but
the second depends on $\vec{m}$ and not $\vec{n}$.  Therefore,
eq.~\eqref{MaudTS} cannot hold, so \textbf{Time Symmetry} is violated.

It is worth pausing for a moment to understand why this model violates
\textbf{Time Symmetry} from a conceptual point of view.  Because the
experiment is its own operational time-reverse, when we exchange the
preparation and measurement variables the quantum predictions can
still be explained by the same ontological model.  This is a sort of
time symmetry and is perhaps what has lead Maudlin astray.  However,
the \textbf{Time Symmetry} assumption requires much more than this.
It says that the ontic description of the time reversed experiment
should be the same as ontic description of the original experiment
running backwards in time.  This means that the way in which the ontic
state is correlated with the preparation in the original experiment
has to be the same as the way in which the ontic state is correlated
with the measurement in the time reversed experiment.  Because the
ontic state contains full information about preparation in the
original experiment, it must contain full information about the
measurement in the time reversed experiment.  This means that the
ontic state must be correlated with the measurement input.  However,
this would require retrocausality, which is not allowed in the
ontological models framework.  Hence, we get a contradiction.  This is
essentially the content of Price's argument for retrocausality.
                                         
\section{Maudlin's Model}

Maudlin's model is just the Beltrametti-Bugajski model specialized to
only two possible preparation and measurement directions, with a
relabelling of the inputs and outputs.

Let the preparation input value $X=0$ correspond to the direction
$\vec{n} = (0,0,1)$ and the value $X = 1$ to the direction
$\vec{n} = (1/2,0,\sqrt{3}/2)$, i.e.\ at $30^{\circ}$ in the $z$-$x$
plane.  Similarly, let the measurement input value $Y = 0$ correspond
to the direction $\vec{m} = (0,0,1)$ and the value $Y = 1$ to the
direction $(-1/2,0,\sqrt{3}/2)$, i.e.\ at $-30^{\circ}$ in the $z$-$x$
plane.  Note the relabelling here as we have chosen $X=1$ and $Y=1$ to
correspond to different directions, whereas $X$ and $Y$ just are the
directions themselves in the Beltrametti-Bugajski model.
Nevertheless, this is allowed as $X$, $Y$, $A$ and $B$ are just
supposed to be abstract labels in our formalism, which can be mapped
to physical degrees of freedom in any way you like.

With these definitions, the operational predictions are
\begin{align*}
  p_{PM}(A = a,B=b|X=0,Y=0) & = \frac{1}{4} \left ( 1 + ab \right ) \\
  p_{PM}(A=a,B=b|X=0,Y=1) & = \frac{1}{4} \left (  1 +
                            \frac{\sqrt{3}}{2}ab \right ) \\ 
  p_{PM}(A = a, B=b|X=1,Y=0) & = \frac{1}{4} \left ( 1 +
                               \frac{\sqrt{3}}{2}ab \right ) \\
  p_{PM}(A=a,B=b|X=1,Y=1) & = \frac{1}{4} \left ( 1 + \frac{1}{2}ab \right ).
\end{align*}
This experiment is its own operational time reverse, i.e.\
$p_{PM}(A=a,B=b|X=x,Y=y) = p_{PM}(A=b,B=a|X=y,Y=x)$.

The reason why Maudlin chose these directions is that these are the
settings used by Bell to show that quantum theory violates local
causality in a spacelike setting, and our main theorem shows that
operationally time reversible experiments that satisfy our assumptions
must satisfy local causality.  Thus, if these predictions could be
reproduced by a model that satisfies our assumptions then that would
be a counter-example to our theorem.

Maudlin's ontological model works in the same way as
Beltrametti-Bugajski, i.e.\ the preparation input $X$ and the output
$A$ are just transmitted to the measurement device.  In this case, the
ontic state space is just $\Lambda = \{0,1\} \times \{+1,-1\}$, so
ontic states are of the form $\lambda = [\lambda_1,\lambda_2]$ with
$\lambda_1 \in \{0,1\}$ and $\lambda_2 \in \{+1,-1\}$.  The
ontological model probabilities are as follows
\begin{align*}
  p_P(A = a|X = x) & = \frac{1}{2} \\
  p_P(L = [\lambda_1,\lambda_2]|X = x,A = \vec{a}) & =
                                                   \delta_{\lambda_1,x} \delta_{\lambda_2,a} \\
  p_M(B = b|L = [0,\lambda_2],Y = 0) & = \frac{1}{2} \left ( 1 +
                                             \lambda_2 b
                                             \right ) \\
  p_M(B = b|L = [0,\lambda_2],Y = 1) & = \frac{1}{2} \left ( 1 +
                                             \frac{\sqrt{3}}{2}\lambda_2 b 
                                               \right ) \\
  p_M(B = b|L = [1,\lambda_2],Y = 0) & = \frac{1}{2} \left ( 1 +
                                             \frac{\sqrt{3}}{2}\lambda_2 b
                                               \right ) \\
  p_M(B = b|L = [1,\lambda_2],Y = 1) & = \frac{1}{2} \left ( 1 +
                                             \frac{1}{2}\lambda_2 b
                                             \right ),
\end{align*}
from which we can calculate
\begin{align*}
  p_{PM}(A=a,B = b,L = [0,\lambda_2]|X=x,Y = 0) & = \frac{1}{4}
                                                  \delta_{0,x}\delta_{\lambda_2,a}
                                                  \left ( 1 +
                                                  \lambda_2 b
                                                  \right )
  \\
  p_{PM}(A=a,B = b,L = [0,\lambda_2]|X=x, Y = 1) & = \frac{1}{4}
                                                   \delta_{0,x}\delta_{\lambda_2,a}
                                                   \left ( 1 +
                                                   \frac{\sqrt{3}}{2}\lambda_2 
                                                   \right ) \\
  p_{PM}(A=a, B = b, L = [1,\lambda_2]|X=x, Y = 0) & = \frac{1}{4}
                                                     \delta_{1,x}\delta_{\lambda_2,a}
                                                     \left ( 1 +
                                                     \frac{\sqrt{3}}{2}\lambda_2 b
                                                     \right ) \\
  p_{PM}(A=a, B = b, L = [1,\lambda_2]|X=x, Y = 1) & = \frac{1}{4}
                                                     \delta_{0,x}\delta_{\lambda_2,a}
                                                     \left ( 1 +
                                                     \frac{1}{2}\lambda_2 b
                                                     \right ). 
\end{align*}

Since the experiment is its own operational time reverse, according to
\textbf{Time Symmetry}, we should have
\begin{equation}
  p_{PM}(A=a,B=b,L=[\lambda_1,\lambda_2]|X=x,Y=y) = p_{PM}(A=b,B=b,L=f[\lambda_1,\lambda_2]|X=x,Y=y).
\end{equation}

If we define $f[\lambda_1,\lambda_2] =
[f_1(\lambda_1),f_2(\lambda_2)]$, $g(a,b,\lambda_1,\lambda_2,x,y) =
p_{PM}(A=a,B=b,L=[\lambda_1,\lambda_2]|X=x,Y=y)$, and
$g(\lambda_1,x,y) =
\sum_{a,b,\lambda_2}g(a,b,\lambda_1,\lambda_2,x,y)$ then, by the same
argument we used for the Beltrametti-Bugajski model we get
\[g(\lambda_1,x,y) = \delta_{\lambda_1,x} =
  \delta_{f_1(\lambda_1,\lambda_2),y},\]
which is again a contradiction as the first expression depends on $x$
but not $y$, and the second depends on $y$ and not $x$.

\section{Lemma VIII.2}

Although we have shown that Maudlin has not provided a counterexample
to our theorem, in his comment he asserts that our Lemma VIII.2 is
incorrect, so we conclude by explaining why he is mistaken.  

In our paper, we give the definition of an ontological time reverse as
in eq.~\eqref{ontTR}, but then immediately assert that if an
ontological time reverse exists then there also exists a time
symmetric ontological model in which $f$ is trivial.  Since Maudlin
claims that our argument for this is difficult to follow, here it is
again in more detail.

The reason we can always find a model in which $f$ is trivial is
because $f$ is invertable.  The condition $L'=f[\lambda]$ is exactly
the same as the condition $f^{-1}[L'] = \lambda$.  We can then define
a new ontic state $L'' = f^{-1}[L]$ and we will have
\begin{equation}
  \label{ontTR2}
  p_{PM}(A = a,B = b,L = \lambda|X = x,Y = y) = p_{P'M'}(A' = b,B' =
  a,L'' = \lambda|X' = y,Y' = x).
\end{equation}
The new ontic state $L''$ inherits all of the conditional
independences that $L$ had because the two are isomorphic.  In
particular, we have the defining equation for an ontological model
\begin{multline}
  \label{OMRev}
  p_{P'M'}(A' = b,B' = a,L'' = \lambda|X' = y,Y' = x) \\ =
  p_{M'}(B' = a|L'' = \lambda,Y = x)p_{P'}(L'' = \lambda|A' = b,X' = y)p_{P'}(A' =
  b|X' = y).
\end{multline}
This holds because the same equation holds if we replace $L''=\lambda$
with $L' = f[\lambda]$, by definition of an ontological model, but 
 $L''=\lambda$ and $L' = f[\lambda]$ are the exact same condition.  

In the paper, we use this construction to prove our results without
having to include the function $f$ in our equations.  In case you do
not buy the above argument, we will now prove the part of Lemma VIII.2
that Maudlin claims is wrong without eliminating $f$.

The relevant claim is that, if an experiment has an ontological time
reverse, than an ontological model of it must satisfy
\begin{equation}
  \label{MI}
  p_{PM}(L=\lambda|X=x,Y=y) = p_P(L=\lambda),
\end{equation}
i.e.\ the ontic state is statistically independent of the measurement
inputs.

To prove this, we start with the defining equation for an ontological
model
\begin{multline}
  p_{PM}(A=a,B=b,L=\lambda|X=x,Y=y) \\ =
  p_{M}(B=b|L=\lambda,Y=y)p_{P}(L=\lambda|X=x,A=a)p_{P}(A=a|X=x).
\end{multline}
We can perform a Bayesian inversion of the second term to obtain
\[p_{P}(L=\lambda|X=x,A=a)p_{P}(A=a|X=x) =
  p_P(A=a|L=\lambda,X=x)p_P(L=\lambda|X=x),\]
which gives
\begin{multline}
  p_{PM}(A=a,B=b,L=\lambda|X=x,Y=y) \\ =
  p_{M}(B=b|L=\lambda,Y=y)p_P(A=a|L=\lambda,X=x)p_P(L=\lambda|X=x).
\end{multline}
Summing both sides over $a$ and $b$ then gives
\begin{equation}
  \label{CI1}
  p_{PM}(L=\lambda|X=x,Y=y) = p_P(L=\lambda|X=x),
\end{equation}
which tells us that, for the experiment $(P,M)$, $L$ is conditionally
independent of $Y$, given $X$.

We now do the same thing for the operational time reverse $(P',M')$,
i.e.\ we start with
\begin{multline}
  p_{P'M'}(A'=b,B'=a,L'=f[\lambda]|X'=y,Y'=x) \\ =
  p_{M'}(B'=a|L'=f[\lambda],Y'=x)p_{P'}(L'=\lambda|X'=y,A'=b)p_{P'}(A'=b|X'=y),
\end{multline}
perform a Bayesian inversion to obtain
\[p_{P'}(L'=f[\lambda]|X'=y,A'=b)p_{P'}(A'=b|X'=y) =
  p_{P'}(A'=b|L'=f[\lambda],X'=y)p_{P'}(L'=f[\lambda]|X'=y),\]
which gives
\begin{multline}
  p_{P'M'}(A'=b,B'=a,L=f[\lambda]|X'=y,Y'=x) \\ =
  p_{M'}(B'=a|L'=f[\lambda],Y'=x)p_{P'}(A'=b|L'=f[\lambda],X'=y)p_{P'}(L'=f[\lambda]|X'=y),  
\end{multline}
and then sum over $a$ and $b$ to obtain
\begin{equation}
  \label{CI2}
  p_{P'M'}(L'=f[\lambda]|X'=y, Y'=x) = p_{P'}(L'=f[\lambda]|X'=y).
\end{equation}

Now, if $(P,M)$ and $(P',M')$ are operational time reverses then we
have
\begin{equation}
  \label{LTR}
  p_{PM}(L=\lambda|X=x,Y=y) = p_{P'M'}(L'=f[\lambda]|X'=y,Y'=x),
\end{equation}
which comes from summing eq.~\eqref{ontTR} over $a$ and $b$.  We
already know that $p_{PM}(L=\lambda|X=x,Y=y) = p_{P}(L=\lambda|X=x)$
from eq.~\eqref{CI1}, so the only question is whether $L$ is also
independent of $X$.  By definition, this is the case if, for every
pair $x_1,x_2\in\Omega_X$, $p_{P}(L=\lambda|X=x_1) =
p_P(L=\lambda|X=x_2)$.  However, applying eqs.~\eqref{CI2} and
\eqref{LTR} we have
\[p_P(L=\lambda|X=x) = p_{P'}(L'=f[\lambda]|X'=y).\]
Since the right hand side does not depend on $x$, the left hand side
must also not vary with $x$, so we have the independence we need.

Maudlin complains that the equation
$p_P(L=\lambda|X=x,Y=y) = p_P(L=\lambda)$ is implausible because it
says that the ontic state is independent of both the preparation and
measurement inputs, so it cannot convey any information at all between
the two, and hence the preparation and measurement variables should be
uncorrelated.  This is not the case for two reasons:

\begin{enumerate}
  \item $p_{P}(L = \lambda)$ can depend on $P$, the choice of preparation
    device.  In quantum theory, for a no-signalling preparation, $P$
    corresponds to a fixed choice of ensemble average density
    operator, i.e.\ $\sum_a \rho_{a|x} = \rho$ is independent of $x$.
    However, different choices of $P$ correspond to different density
    operators $\rho$, and $p_{P}(L = \lambda)$ can depend on this.
    Admittedly, this is somewhat obscured by the fact that we dropped
    the subscripts $_{PM}$ in the main argument of our paper.
  \item Just because $L$ is independent of $X$ does not mean that it
    is independent of $X$ and $A$, taken together, i.e.\
    $p_{P}(L= \lambda|X=x,A=a) \neq p_P(L=\lambda)$.  All we require
    is that
    $\sum_{a\in \Omega_A} p_P(L=\lambda|X=x,A=a)p_P(A=a|X=x) =
    p_P(\lambda)$.  Thus, $L$ can transmit information about $A$ and
    $X$ to the measurement device, provided it does so in such a way
    that there is no information about $X$ left over when we average
    over $A$.  If $P$ is a no-signalling preparation, then this is
    exactly Spekkens' criterion of preparation noncontextuality
    \cite{Spekkens2005}, as we explained in our paper.  The example of
    Spekkens' toy theory, also explained in our paper, shows how one
    can get nontrivial correlations between the preparation and
    measurement variables using this mechanism.
\end{enumerate}

\bibliographystyle{unsrturl}
\bibliography{sym}

\end{document}